\magnification=1210
\baselineskip=15pt
\def\medskip{\vskip .2in}

\def\ds{\displaystyle}
\pageno=1
\footline{\hss\folio\hss}
\vskip2.5pc
\centerline{\bf  Effects on the eigenvalues of the quantum
bouncer due to dissipation}
\vskip2pc
\centerline{ G. L\'opez and  G. Gonz\'alez}
\vskip1pc
\centerline{Departamento de F\'isica de la Universidad de Guadalajara}
\centerline{Apartado Postal 4-137}
\centerline{44410~Guadalajara, Jalisco, M\'exico}
\vskip5pc
\centerline{Nov 2003} 
\vskip1pc
\centerline{PACS~~03.20.+i~~~03.65.Ca}
\centerline{keywords:~~dissipation, constant of motion, quantum bouncer}

\vskip4pc
\centerline{ABSTRACT}
\vskip2pc
Effects on the spectra of the quantum bouncer due to dissipation are given when a linear or
quadratic dissipation is taken into account. Classical constant of motions and Hamiltonians
are deduced for these systems and their quantized eigenvalues are estimated 
through perturbation theory. we found some differences when we compare the eigenvalues of
these two quantities.
\vfil\eject
\leftline{\bf I. Introduction.}
\vskip0.5pc
Dissipative systems has been one  of the must subtle and difficult topics to deal with in
classical [1] and quantum physics [2]. In general, to construct a consistent Lagrangian and
Hamiltonian formulation for a given dissipative system can be a really challenge [3]. There
are basically two approaches to study dissipative systems. The first one tries to bring
about the dissipation as a results of averaging over all the coordinates of the batch
system, where one considers the whole system as composed of two parts, our original
conservative system and the batch system which interacts 
with the conservative system and causes the dissipation (of energy) on it [4]. This approach has
its own value and will not be followed or discussed here. The second approach considers
that the bath system produces on our initially conservative system and average effect which 
is expressed as an additional external velocity depending force acting on the conservative
system and transforming it into a dissipative system with this velocity depending force, the
resulting classical dissipative system contains then this phenomenological (or theoretical)
velocity depending force. Then, the question arises over its consistent Lagrangian and
Hamiltonian formalism and the consequences of its quantization. This approach, in addition,
allows us to study and test the Hamiltonian approach for quantum mechanics and its
consistence [5] and is the approach we will follow in this paper. A system which has
bring our attention for dissipation study through the above approach is the quantum
bouncer. The quantum bouncer [6] is the quantization of the motion of a
particle which is attracted by the constant gravity force, that is, close to surface of
the earth. This particle hits a perfectly reflexing surface, producing the bouncing
effect. This system with an additional dissipation force has particular importance because
of its potential experimental realization. This dissipative system has been
studied very little and using so far the first approach mentioned above [7].
\vskip0.5pc\noindent
In this paper we considers the second approach to
dissipation systems and will assume that the external velocity depending force has 
linear and quadratic dependency with respect the velocity. This approach gives us the
opportunity to check the nature of quantization via Hamiltonian or constant of motion
associated to the system, that is, using the usual quantization of the linear generalized
momentum or using the quantization of the velocity. This consideration is particularly
interesting  in dissipative systems since one can not always have a Hamiltonian in an
explicit form as a function of the variables position and linear momentum [8], that is,
the velocity "v" can not always be known explicitly in terms of the linear momentum "p"
and position "x" of the particle through the relation $p=\partial L/\partial v$, where
$L$ is the Lagrangian of the system.  This paper is
organized as follows: we present the classical study for the dissipative system considering
the linear and quadratic velocity depending force. The constant of motion, the Lagrangian,
and the Hamiltonian of the system are derived, and we give their expressions up to second
order in the dissipation parameter. We presents the modification of the eigenvalues for the
quantum bouncer, when this dissipation is taken into account, for the above approximated
(weak dissipation) constant of motion and Hamiltonian using 
quantum perturbation theory. Finally, conclusions and some discussions of our results are
made.
\vskip1pc
\leftline{\bf II. Classical linear dissipation. }
\vskip0.5pc\noindent
The motion of a particle of mass $m$ under a constant gravitational force and a linear
dissipative force is described by the equation 
$$m{d^2x\over dt^2}=-mg-\alpha v\ ,\eqno(1)$$
where $x$ is the position of the particle, $g$ is the constant acceleration due to earth
gravity, $\alpha$ is the parameter which characterizes the dissipation, and $v=dx/dt$ is
the velocity of the particle. A constant of motion of the autonomous system (1) is a
function $K_{\alpha}=K_{\alpha}(x,v)$ satisfying the equation [9]
$$v{\partial K_{\alpha}\over\partial x}-(g+{\alpha\over m}v){\partial K_{\alpha}\over v}=0\
.\eqno(2)$$
The solution of this equation such that $\lim_{\alpha\to 0}K_{\alpha}=mv^2/2+mgx$ (the
usual total energy of the non dissipative system) is given by
$$K_{\alpha}={m^2gv\over\alpha}-m\left({mg\over\alpha}\right)^2\ln(1+{\alpha v\over mg})+mgx
\ .\eqno(3)$$
The Lagrangian associated to (1) can be obtained using the known expression [5]
$$L_{\alpha}=v\int{K_{\alpha}(x,v)\over v^2}dv\ ,\eqno(4)$$
bringing about the following Lagrangian
$$L_{\alpha}={m^2gv\over\alpha}\ln\left(1+{\alpha v\over
mg}\right)+m\left({mg\over\alpha}\right)^2
\ln\left(1+{\alpha v\over mg}\right)-mgx -{m^2gv\over\alpha}\ .\eqno(5)$$
Therefore, the generalized linear momentum and Hamiltonian are given by
$$p_{\alpha}={m^2g\over\alpha}\ln\left(1+{\alpha v\over mg}\right)\eqno(6)$$
and
$$H_{\alpha}=m\left({mg\over\alpha}\right)^2\left(\exp\left({\alpha p_{\alpha}\over m^2
g}\right)-1\right)-{mg\over\alpha}p_{\alpha}+mgx\ .\eqno(7)$$
At two orders in the dissipation parameter $\alpha$, one has the constant of motion, the
Lagrangian, the generalized linear momentum, and Hamiltonian given as
$$K={1\over 2}mv^2+mgx-{\alpha\over 3g}v^3+{\alpha^2\over 4mg^2}v^4\ ,\eqno(8a)$$
$$L={1\over 2}mv^2-mgx-{\alpha\over 6g}v^3+{\alpha^2\over 12mg^2}v^4\ ,\eqno(8b)$$
$$p =mv-{\alpha\over 2g}v^2+{\alpha^2\over 3mg}v^3\ ,\eqno(8c)$$
and
$$H={p^2\over 2m}+mgx+{\alpha\over 6mg}p^3+{\alpha^2\over 24m^5g^2}p^4\ .\eqno(8d)$$
The constant  of motion (3)  or (8a)  and the
Hamiltonian (7) or (8d) bring about the damping bouncing effect on the spaces ($x,v$) and
($x,p$). The dissipative parameter $\alpha$ can be determined by measuring the velocity
$v_o$ at the reflexing surface ($x=0$) and then measuring its maximum displacement
$x_{max}$ ($v=0$). Equaling the value of the constant of motion on both situations, one gets
the expression
$${m^2gv_o\over\alpha}-m\left({mg\over\alpha}\right)^2\ln(1+{\alpha v_o\over
mg})=mgx_{max}\ , \eqno(9)$$
from the parameter $\alpha$ can be gotten.
\vskip0.5pc\noindent
{\bf III. Classical quadratic dissipation.}
\vskip0.5pc\noindent
In this case, the motion of the particle is described by the equation
$$m{d^2x\over dt^2}=-mg-\gamma v|v|\ ,\eqno(10)$$
where $\gamma$ represents a dissipation constant which, of course, is different from the
previous case. Proceeding in the same way as we did for the linear case, the constant of
motion, Lagrangian, generalized linear momentum, and Hamiltonian are given by
$$K_{\pm}={1\over 2}mv^2\exp\left(\pm{2\gamma x\over m}\right)\pm {m^2 g\over
2\gamma}\left(
\exp\left(\pm{2\gamma x\over m}\right)-1\right)\ ,\eqno(11a)$$
$$L_{\pm}={1\over 2}mv^2\exp\left(\pm{2\gamma x\over m}\right)\mp {m^2 g\over
2\gamma}\left(
\exp\left(\pm{2\gamma x\over m}\right)-1\right)\ ,\eqno(11b)$$
$$p_{\pm}=mv\exp\left(\pm{2\gamma x\over m}\right)\ ,\eqno(11c)$$
and
$$H_{\pm}={p_{\pm}^2\over 2 m}\exp\left(\mp{2\gamma x\over m}\right)\pm {m^2 g\over
2\gamma}\left( \exp\left(\pm{2\gamma x\over m}\right)-1\right)\ ,\eqno(11d)$$ 
where the upper sign corresponds to the case $v\ge 0$, and the lower sign corresponds to
the case $v<0$. These equation where already given in reference [10]. The damping effect
of the bouncing particle in the space ($x,v$) can be traced in the following way: starting
with the initial condition $x_o=0$ and $v_o>0$, for example, the constant of motion $K_{+}$
is determined, $K_{+}=mv_o^2/2$. Then, the maximum distance $x_{max}$ ($v=0$) is calculated
from the expression $K_{+}=(m^2g/2\gamma)(exp(2\gamma x_{max}/m)-1)$ which helps to
calculate the constant $K_{-}$, $K_{-}=-(m^2g/2\gamma)\bigl(\exp(-2\gamma
x_{max}/m)-1\bigr)$. This
$K_{-}$ is used now to calculated the velocity at the return point ($x=0$),
$v_1^*=-\sqrt{2K_{-}/m}$. Due to perfectly reflexing surface, the velocity of the
bouncing particle for the next cycle is
$v_1=-v_1^*$ ($v_1<v_o$), and the above cycle is reproduced again, and so on. Starting
with the same initial conditions, the trajectories in this space are one below the other
at any time, as the damping factor is grater. The damping effect in the space ($x,p$)
through the Hamiltonian approach can be analyzed similarly. However, the trajectories
starting with the same initial conditions on this space are not below the other all the
time, as the damping factor is greater. This strange effect is due to change in sign in
(11d) with respect to (11a), produced by the position and velocity dependence of the
expression (11c). 
\vskip0.5pc
To determine the constant $\gamma$ through the constant of motion, one can start with the
initial conditions ($x_o=0,v_o>0$) and can determined the constant of motion
$K_{+}=mv_o^2/2$. Then, one can measure the maximum displacement $x_{max}$ ($v=0$) and
to solve $\gamma$ from the equation
$${1\over 2}mv_o^2={m^2g\over 2\gamma}\left(\exp\left({2\gamma x_{max}\over
m}\right)-1\right)\ .\eqno(12)$$
Up to second order in the dissipation parameter, one has from (11a) to (11d) the constant
of motion, the Lagrangian, the generalized linear momentum, and the Hamiltonian given by
$$K_{\pm}={1\over 2}mv^2+mgx\pm\gamma[v^2x+gx^2]+\gamma^2[v^2x^2/m+2gx^3/3m]\ ,\eqno(13a)$$
$$L_{\pm}={1\over 2}mv^2-mgx\pm\gamma[v^2x-gx^2]+\gamma^2[v^2x^2/m-2gx^3/3m]\ ,\eqno(13b)$$
$$p_{\pm}=mv\pm\gamma[2vx]+\gamma^2[2vx^2/m]\ ,\eqno(13c)$$
$$H_{\pm}={p^2\over 2m}+mgx\mp\gamma[p^2x/m^2- gx^2]+\gamma^2[p^2x^2/m^3+2gx^3/3m]\
.\eqno(13d)$$ 
\vskip1pc
\leftline{\bf IV. Quantization of the constant of motion.}
\vskip0.5pc\noindent
Eqs. (8a) and (13a) can be written as
$$K(x,v)=K_o(x,v)+V(x,v)\ ,\eqno(14)$$
where $K_o$ is the constant of motion without dissipation
$$K_o(x,v)={1\over 2}mv^2+mgx\ ,\eqno(15a)$$
and $V$ takes into account the dissipation factors
$$V(x,v)=\cases{-\alpha\left({\ds v^3\over\ds 3g}\right)+\alpha^2\left({\ds v^4\over\ds
4mg^2}\right)
\quad\quad\quad\quad\quad\quad\quad\hbox{(linear
case)}\cr\cr
\mp\gamma\left[ v^2x+ gx^2\right]+\gamma^2\left[{\ds v^2x^2\over\ds m}+
{\ds 2gx^3\over\ds 3m}\right]\quad\hbox{(quadratic case)}}\eqno(15b)$$
The quantization of (14) can be carried out through the associated Schr\"odinger's equation
of this constant of motion
$$i\hbar{\partial\Psi\over\partial t}=\widehat K(\widehat x,\widehat v)\Psi\ ,\eqno(16)$$
where $\Psi=\Psi(x,t)$ is the wave function, $\hbar$ is the Plank constant divided by
$2\pi$, $\widehat K=\widehat K_o+\widehat V$ is a Hermitian operator associated to (17),
and $\widehat v$ is the velocity operator defined as 
$$\widehat v=-{i\hbar\over m}{\partial\over\partial x}\ .\eqno(17)$$
Since Eq. (16) represents an stationary problem, the usual proposition\break
$\Psi(x,t)=\exp(-iE^Kt/\hbar)~\psi(x)$ transforms (16) to an eigenvalue problem
$$(\widehat K_o+\widehat V)\psi=E^K\psi\ .\eqno(18)$$
Considering now the operator $\widehat V$ as a perturbation of the constant of motion
$K_o$, one can calculate an approximated solution to the
problem (18) through perturbation theory .  The solution of the eigenvalue problem
$$\widehat K_o\psi_n^{(0)}=E_n^{(0)}\psi_n^{(0)}\eqno(19)$$
is well known [6], with $\psi_n^{(0)}$ being the eigenfunction  given by
$$\psi_n^{(0)}={Ai(z-z_n)\over |Ai'(-z_n)|}\ ,\eqno(20)$$
where $Ai$ and $Ai'$ are the Airy function and its differentiation, and $z_n$ is its
nth-zero ($Ai(-z_n)=0$) which ocurres for negative argument only.  $z$ is the normalized
variable
$z=x/l_g$ with
$l_g=\left(\hbar^2/2m^2g\right)^{1/3}$, and $z_n$ is related with the eigenvalue $E_n^{(0}$
through the expression
$$z_n={E_n^{(0)}\over mgl_g}\ .\eqno(21)$$
Up to second order in perturbation theory, the eigenvalues of (18) are given (in Dirac
notation [11]) as
$$E_n^K=E_n^{(0)}+\langle n|\widehat V|n\rangle+
\sum_{k\not= n}{|\langle n|\widehat V|k\rangle|^2\over E_k^{(0)}-E_n^{(0)}}\ ,\eqno(22)$$
where $\langle z|n\rangle=\psi_n^{(0)}$. Using the Hermitian operators 
$\widehat {v^2x}=({\widehat v}^2x+{\widehat v} x {\widehat v}+x {\widehat v}^2)/3$ and
$\widehat{v^2x^2}=({\widehat v}^2x^2+{\widehat v}x^2{\widehat v}+x^2{\widehat v}^2+
x{\widehat v}^2x+x{\widehat v}x{\widehat v}+{\widehat v}x{\widehat v}x)/6$  
for the associated expressions on (25b), and using the relations $\langle
n|x^s|k\rangle=l_g^s\langle n|z^s|k\rangle$ and
$\langle n|d^s/dx^s|k\rangle=l_g^{-s}\langle n|d^s/dz^s|k\rangle$ for any integer $s$, 
one has (see appendix for a list of matrix elements)
$$E_n^K=E_n^{(0)}+\cases{
\alpha^2\left[{\ds l_g^2z_n^2\over\ds 5m}+{\ds 8\over\ds 9}gl_g^3
\ds{\sum_{k\not=n}}{\ds |1/2+mgl_g/(E_k^{(0)}-E_n^{(0)})|^2\over\ds
E_k^{(0)}-E_n^{(0)}}\right]
\quad\quad\quad\quad\hbox{(linear)}
\cr\cr\cr
\mp ~\gamma{\ds 12 g l_g^2z_n^2\over\ds 15}+
\gamma^2\left[\left(-{\ds 1\over\ds 2}+{\ds 56 z_n^3\over\ds 105}\right){\ds
2gl_g^3\over\ds m} +4g^2l_g^4{\ds\sum_{k\not=n}}a_{nk}
\right]\ ,\quad\hbox{(quadratic)}\cr}\eqno(23a)$$
where $a_{nk}$ are real number given by
$$a_{nk}={\ds | 12-2z_k(z_n-z_k)^2+(z_n-z_k)^3|^2\over\ds (z_k-z_n)^9}\ .\eqno(23b)$$
Note that for the linear dissipation case, there  is not real contribution a first
approximation, and for the quadratic dissipative case, the first order contribution
depends on whether the particle is moving (-) or down (-). Within a full cycle, this
first order is cancelled out and the second order contribution remains. Of course, for
the approximation (3a) to be valid, one must have that the second term in this expression
must be much less than $E_n^{(0)}$ which restricts the value of the dissipative parameter.  
\vfil\eject
\leftline{\bf V. Quantization of the Hamiltonian.}
\vskip0.5pc\noindent
Eqs. (8d) and (13d) can be written as
$$H(x,p)=H_o(x,p)+W(x,p)\ ,\eqno(24)$$
where $H_o$ is the Hamiltonian without dissipation,
$$H_o(x,p)={p^2\over 2m}+mgx\ ,\eqno(25a)$$
and $W$ has the dissipation terms,
$$W(x,p)=\cases{\alpha\left({\ds p^3\over\ds 6mg}\right)+\alpha^2\left({\ds p^4\over\ds
24m^5g^2}\right)\quad\quad\quad\quad\hbox{(linear)}\cr\cr
\mp\gamma[{\ds p^2x\over\ds m^2}- gx^2]+\gamma^2[{\ds p^2x^2\over\ds m^3}+{\ds 2gx^3\over
\ds 3m}]\quad\hbox{ (quadratic)}}\eqno(25b)$$
It is necessary to mention that the quantization of some systems for quadratic
dissipation has been solved by different authors [10,12] but perfectly
reflexing wall potential,
$$\tilde V(x)=\cases{\infty&for $x<0$\cr\cr
mgx&for $x\ge 0$\cr}\ .\eqno(26)$$
 Moreover, the solution given in reference [10] is singular
for the dissipation parameter equal to zero. Therefore, we think it worths to make the
analysis of the quantization for small orders in the parameter $\gamma$. For the usual
Shr\"odinger quantization approach, one has the stationary equation
$$i\hbar{\partial\Psi\over\partial t}=\widehat H(x,\widehat p)\Psi\ ,\eqno(27)$$
where $\widehat H$ is the Hamiltonian operator associated to (24), and $\widehat p$ is the
usual linear momentum operator $\widehat p=-i\hbar{\partial/\partial x}$. Eq. (27) is
transformed to an eigenvalue problem, $\widehat H\psi(x)=E^H\psi(x)$, through the
proposition $\Psi(x,t)=\exp\left(-iE^H t/\hbar\right)\psi(x)$. Since the Hamiltonian
$\widehat H$ is given by $\widehat H=\widehat H_o+\widehat W$, where the solution of the
equation 
$$\widehat H_o\psi_n^{(0)}=E_n^{(0)}\psi_n^{(0)}\eqno(28)$$
is well known (as before), being $\psi_n^{(0)}$ and $E_n^{(0)}$  given by (20) and (21),
perturbation theory can be used to determine the approximated values of the eigenvalues
$E_n^H$ (similarly as done with expression (22)). Using the Hermitian operators
$\widehat{p^2x}=({\widehat p}^2x+{\widehat p}x{\widehat p}+x{\widehat p}^2)/3$ and
$\widehat{p^2x^2}=({\widehat p}^2x^2+{\widehat p}x^2{\widehat v}+x^2{\widehat p}^2+
x{\widehat p}^2x+x{\widehat p}x{\widehat p}+{\widehat p}x{\widehat p}x)/6$  
for the associated expressions on (25b), one gets
$$E_n^H=E_n^{(0)}+\cases{
\alpha^2\left[{\ds l_g^2z_n^2\over\ds 30m}+{\ds 4\over\ds 9}gl_g^3
\ds{\sum_{k\not=n}}{\ds |1/2+mgl_g/(E_k^{(0)}-E_n^{(0)})|^2\over\ds
E_k^{(0)}-E_n^{(0)}}\right]
\quad\hbox{(linear)}
\cr\cr\cr
\pm~ \gamma{\ds 4 g l_g^2z_n^2\over\ds 15}+
\gamma^2\left[\left(-{\ds 1\over\ds 2}+{\ds 56 z_n^3\over\ds 105}\right){\ds
2gl_g^3\over\ds m} +4g^2l_g^4{\ds\sum_{k\not=n}}a_{nk}
\right]\ ,\quad\hbox{(quadratic)}\cr}\eqno(29)$$
where $a_{nk}$ is given by (23b).
As one can see from (23) and (29), there is a difference with the eigenvalues associated
to the constant of motion and Hamiltonian quantizations. Their relative differences,
$\delta E_n=(E_n^H-E_n^K)/E_n^{(0)}$, is given by
$${\delta E_n\over E_n^{(0)}}=\cases{
\alpha^2\left[-{\ds 1\over\ds 6}{\ds l_gz_n\over\ds 6 m^2g}-{\ds 4\over\ds 9}{\ds l_g^2
\over\ds m z_n}
\ds{\sum_{k\not=n}}{\ds |1/2+mgl_g/(E_k^{(0)}-E_n^{(0)})|^2\over\ds
E_k^{(0)}-E_n^{(0)}}\right]
\quad\hbox{(linear)}
\cr\cr\cr
\pm~ \gamma{\ds 16 g l_g^2z_n^2\over\ds 15}\ ,\hskip15pc\hbox{(quadratic)}\cr}\eqno(30)$$

\vskip1pc
\leftline{\bf VI. Conclusion.}
\vskip0.5pc\noindent
We have presented the classical and quantum analysis of a particle attracted by constant
gravity and velocity depending dissipative forces which bounce on a perfectly reflexing
surface. We have considered linear and quadratic velocity depending dissipative cases and
have deduced their constant of motion and Hamiltonians. Expression (3) and (11a) gives us 
the expected damping behavior of the particle on the space ($x,v$), but the expression (7)
and (11d) show us the damping with unexpected behavior in the space $(x,p$)
(two trajectories on this space, for dissipative parameter one bigger than other, do
not follow one under the other all the time).    For the quantum case, we have
analyzed the eigenvalues of the assigned operators pertubedly and up to second order 
on the dissipative parameters. Relation (30) tells us that that there is a difference
whether the constant of motion  or the Hamiltonian is quantized, and it suggests that
one could see this difference experimentally. In this way, one could see whether nature
prefers to follow constant of motions rather than Hamiltonians for dissipative systems.
 Finally, one must observe that for the full linear case (7), it is possible to solve exactly
the Shr\"odinger equation in the momentum representation, and this will be analyzed on a future
paper.
\vfil\eject
\leftline{APPENDIX}
\vskip1pc\noindent
We show here a list of some matrix elements from reference [6] and some other calculated
from the same reference (an correction of a sign has been made to some matrix elements).
Given the functions (20) and
$n\not=k$, one has
$$\eqalign{
\langle n|k\rangle &=\delta_{nk}\hskip24pc(A_1)\cr\cr
\langle n|z| n\rangle &={2\over 3}z_n\hskip5pc \langle n|z|k\rangle ={\ds
2(-1)^{n+k+1}\over (z_n-z_k)^2}\hskip11pc (A_2)\cr\cr
\langle n|z^2|n\rangle &={8\over 15}z_n^2\hskip5pc \langle n|z^2|k\rangle =
{\ds 24(-1)^{n+k+1}\over (z_n-z_k)^4}\hskip9.5pc(A_3)\cr\cr
\langle n|z^3|n\rangle &={3\over 7}+{48\over 105}z_n^3\hskip3pc
\langle n|z^3|k\rangle={\ds 24(z_n+z_k)(-1)^{n+k+1}\over (z_n-z_k)^4}\hskip6pc(A4)\cr\cr
\langle n|{\ds d\over dz}|n\rangle &=0\hskip6.5pc
\langle n|{\ds d\over dz}|k\rangle={\ds (-1)^{n+k}\over z_n-z_k}\hskip11pc(A_5)\cr\cr
\langle n|{\ds d^2\over dz^2}|n\rangle &=-{1\over 3} z_n\hskip4.8pc
\langle n|{\ds d^2\over dz^2}|k\rangle={\ds 2 (-1)^{n+k}\over (z_n-z_k)^2}
\hskip10pc(A_6)\cr\cr
\langle n|{\ds d^3\over dz^3}|n\rangle &={1\over 2}\hskip6pc
\langle n|{\ds d^3\over dz^3}|k\rangle=\left({1\over 2}+{1\over z_k-z_n}\right)(-1)^{n+k}
\hskip5pc(A_7)\cr\cr
\langle n|{\ds d^4\over dz^4}|n\rangle &={1\over 5}z_n^2\hskip5pc
\langle n|{\ds d^4\over dz^4}|k\rangle={\ds -2(z_k-z_n)+24-2z_k(z_k-z_n)^2\over
(z_k-z_n)^4}(-1)^{n+k}\cr
&\hskip26pc(A_8)
}$$

\vfil\eject
\leftline{\bf References}
\vskip0.5pc\noindent
\obeylines{
[1] R. Glauber and V.I. Man'ko, Sov. Phys. JEPT 60 (1984)450.
\quad V.V. Dodonov, Hadron J.,{\bf 4} (1981) 173.
\quad G. L\'opez, M. Murg\'{\i}a and M. Sosa, Mod. Phys. Lett. B, {\bf 11}, 14 (1997) 625.
[2] M. Razavy, Can. J. Phys. {\bf 50} (1972) 2037.
\quad S. Okubo, Phys. Rev. A, {\bf 23} (1981) 2776.
\quad G. L\'opez, M. Murg\'{\i}a and M. Sosa, Mod. Phys. Lett. B, {\bf 15}, 22 (2001) 965.
[3] M. Mijatovic, B. Veljanoski and D. Hajdukovic, Hadronic J. {\bf 7}, 5 (1984) 1207.
\quad G. L\'opez, Ann. Phys., {\bf 251},2 (1996) 372.
\quad G. L\'opez and G. Gonz\'alez, IL Nuo. Cim. B, {\bf 118},2 (2003) 107.
[4] A.O. Caldeira and A.T. Leggett, Physica A, {\bf 121} (1983) 587.
\quad W.G. Unruh and W.H. Zurek, phys. Rev. D, {\bf 40} (1989) 1071.
\quad B.L. Hu, J.P. Paz and Y. Zhang, Phys. Rev. D, {\bf 45} (1992) 2843.
\quad G.P. Berman, F. Borgonovi, G.V. L\'opez and V.I. Tsifrinovich, 
\quad Phys. Rev. A, {\bf 68} (2003) 012102.
[5] G. L\'opez, Rev. Mex. Fis., {\bf 48} ,1 (2002) 10.
[6] J. Gean-Banacloche, Am. J. Phys. {\bf 67}, 9 (1999) 776.
\quad D.M. Goodmanson, Am. J. Phys. {\bf 68}, 9 (2000) 866.
[7] R. Onofrio and L. Viola, quant-ph/9606024 v1 (1996).
[8] G. L\'opez, Rev. Mex. Fis.,{\bf 45} (1999) 1817.
[9] G. L\'opez, "{\it Partial Differential Equations of First Order and Their Applications
\quad to Physics}", World Scientific (1999) page 37.
[10] F. Negro and A. Tartaglia, Phys. Lett. A, {\bf 77} (1980) 1.
\quad F. Negro and A. Tartaglia, Phys. Rev. A, {\bf 1981} (23) 1591.  
[11] P.A.M. Dirac, "{\it The Principles of Quantum Mechanics}",
\quad Oxford Science Publications, 1992.
[12] J.S. Borges, L.N. Epele and H. Fanchiotti, Phys. Rev. A, {\bf 38}, 6 (1988) 3101.
\quad C. Stuckeno and D.H. Kobe, Phys. Rev. A, {\bf 34}, 5 (1986) 3565.
\quad M. Mijatovic, B. Veljanoski and D. Hajdukovic, Hadronic J.,{\bf 7}, 5(1984) 1207.
\quad B.W. Huang, Z.V. Gu and S.W. Qiam, Phys. Lett A, {\bf 142}, 4-5 (1989) 203.
\quad M. Razavy, Hadronic J., {\bf 6},2 (1983) 406.
 }
\end